\documentclass[fleqn,10pt]{wlscirep}
\title{Transport evidence for a sliding two-dimensional quantum electron solid}
\author[1]{Pedro Brussarski}
\author[2]{S. Li}
\author[1,3,*]{S.~V. Kravchenko}
\author[4]{A.~A. Shashkin}
\author[2]{M.~P. Sarachik}

\affil[1]{Physics Department, Northeastern University, Boston, Massachusetts 02115, USA}
\affil[2]{Physics Department, City College of the City University of New York, New York, New York 10031, USA}
\affil[3]{Institute of Fundamental and Frontier Sciences, University of Electronic Science and Technology of China, Chengdu 610054, People's Republic of China}
\affil[4]{Institute of Solid State Physics, Chernogolovka, Moscow District 142432, Russia}
\affil[*]{s.kravchenko@northeastern.edu}

\begin{abstract}
Ignited by the discovery of the metal-insulator transition, the behaviour of low-disorder two-dimensional (2D) electron systems is currently the focus of a great deal of attention. In the strongly-interacting limit, electrons are expected to crystallize into a quantum Wigner crystal, but no definitive evidence for this effect has been obtained despite much experimental effort over the years. Here, studying the insulating state of a 2D electron system in silicon, we have found two-threshold voltage-current characteristics with a dramatic increase in noise between the two threshold voltages. This behaviour cannot be described within existing traditional models. On the other hand, it is strikingly similar to that observed for the collective depinning of the vortex lattice in Type-II superconductors. Adapting the model used for vortexes to the case of an electron solid yields good agreement with our experimental results, favouring the quantum electron solid as the origin of the low-density state.
\end{abstract}
\begin{document}
\flushbottom
\maketitle

\thispagestyle{empty}
\section*{Introduction}
Recently, there has been progress in understanding the metallic side of the metal-insulator transition in low-disorder strongly-interacting 2D electron systems\cite{abrahams01,kravchenko04,shashkin05,punnoose2005,anissimova2007,spivak2010,camjayi08,qiu2012,pudalov2012,dolgopolov2017}, while the origin of the low-density state on the insulating side has remained a mystery. Experimental investigations of the transport and thermodynamic properties of 2D electrons in semiconductors have suggested that these systems approach a phase transition at low electron densities to a new, unknown state that could be a quantum Wigner crystal or a precursor\cite{wigner34,chaplik72,tanatar89,attaccalite02,mokashi12,melnikov2017}; the term quantum means that the corresponding kinetic energy of 2D electrons is determined by the Fermi energy in contrast to the well-known classical Wigner crystal\cite{grimes1979} in which the kinetic energy of electrons is determined by temperature. The phase transition point in the least-disordered 2D electron systems in semiconductors was found to be close to the critical electron density for the metal-insulator transition, below which the 2D electrons become localized (immobile). Although the insulating side of the metal-insulator transition has been extensively studied\cite{andrei88,pudalov93,knighton2018}, no definitive conclusion has been reached concerning the origin of the low-density state. Nonlinear current-voltage ($I-V$) curves were observed and interpreted as either manifestation of the depinning of an electron solid\cite{pudalov93,williams91,chitra05} or the breakdown of the insulating phase within traditional scenarios such as strong electric field Efros-Shklovskii variable range hopping\cite{shklovskii92} or percolation (see, \textit{e.g.}, Refs.\cite{jiang1991,shashkin05} and references therein). The observation of broad-band voltage noise at the threshold $I-V$ curves as well as the attempt to probe the low-density state in perpendicular magnetic fields also have not provided information that allows a choice between the depinning of the electron solid or traditional mechanisms\cite{jiang1991,shashkin05}. It is worth noting that much confusion was introduced by the fact that many authors chose to interpret their data in terms of Wigner crystal, ignoring mundane interpretations.

Here we report a significant breakthrough in our understanding of the origin of the low-density state in a strongly interacting 2D electron system in silicon metal-oxide-semiconductor field-effect transistors (MOSFETs). We have observed two-threshold $V-I$ characteristics with a dramatic increase in noise between the two threshold voltages at the breakdown of the insulating state. In the form of fluctuations with time, the noise in current increases dramatically above $V_{\text{th1}}$ and essentially disappears above $V_{\text{th2}}$. It is the sharp noise peak on the $V-I$ curves that makes the two-threshold behaviour evident. The double threshold behaviour is very similar to that observed for the collective depinning of the vortex lattice in Type-II superconductors (see, \textit{e.g.}, Ref.\cite{blatter94}) provided the voltage and current axes are interchanged. This strongly favours the sliding 2D quantum electron solid whereas the double threshold behaviour cannot be described within alternative scenarios such as percolation or overheating. We emphasize that rather than being an ideal Wigner crystal, the 2D electron system under study is likely to be closer to an amorphous solid, which is similar to the case of the vortex lattice in Type-II superconductors where the collective pinning was observed. It is important to note that in the earlier studies, where current was passed through the sample and the voltage between potential probes was measured, no distinct features were found on almost flat $V(I)$ curves in the breakdown regime. The main problem with those studies is that the noise in the voltage signal is small, which prevents one from identifying a second threshold voltage. By contrast, we applied dc voltage between source and drain and measured the induced current. We found that in this measurement configuration, the noise in current is clear and dramatic. It is this novel experimental protocol that enabled the discovery of features that had not been observed before.

\begin{figure}\hspace*{-3mm}
\scalebox{0.8}{\includegraphics{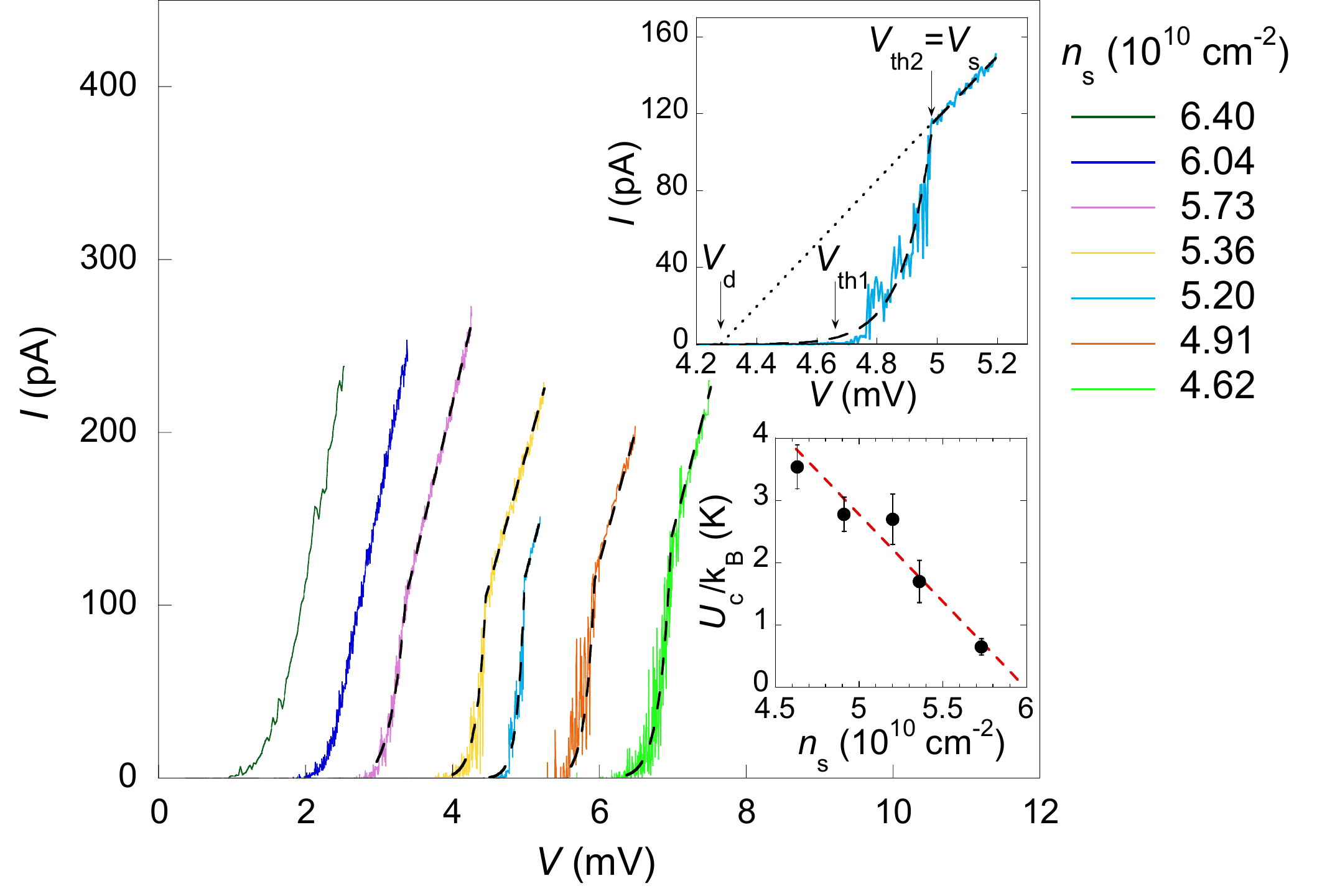}}
\caption{\label{fig1} Voltage-current characteristics. $V-I$ curves are shown for different electron densities in the insulating state at a temperature of 60~mK. The dashed lines are fits to the data using Eq.~(\ref{I}). The top inset shows the $V-I$ curve for $n_{\text{s}}=5.20\times 10^{10}$~cm$^{-2}$ on an expanded scale; also shown are the threshold voltages $V_{\text{th1}}$ and $V_{\text{th2}}$, the static threshold $V_{\text{s}}=V_{\text{th2}}$, and the dynamic threshold $V_{\text{d}}$ that is obtained by the extrapolation of the linear region of the $V-I$ curve to zero current. Bottom inset: activation energy $U_{\text{c}}$ \textsl{vs}.\ electron density.  Vertical error bars represent standard deviations in the determination of $U_{\text{c}}$ from the fits to the data using Eq.~(\ref{I}).  The dashed line is a linear fit.}
\end{figure}

\section*{Results}
\subsection*{Voltage-current characteristics.}

\begin{figure}
\scalebox{0.7}{\includegraphics{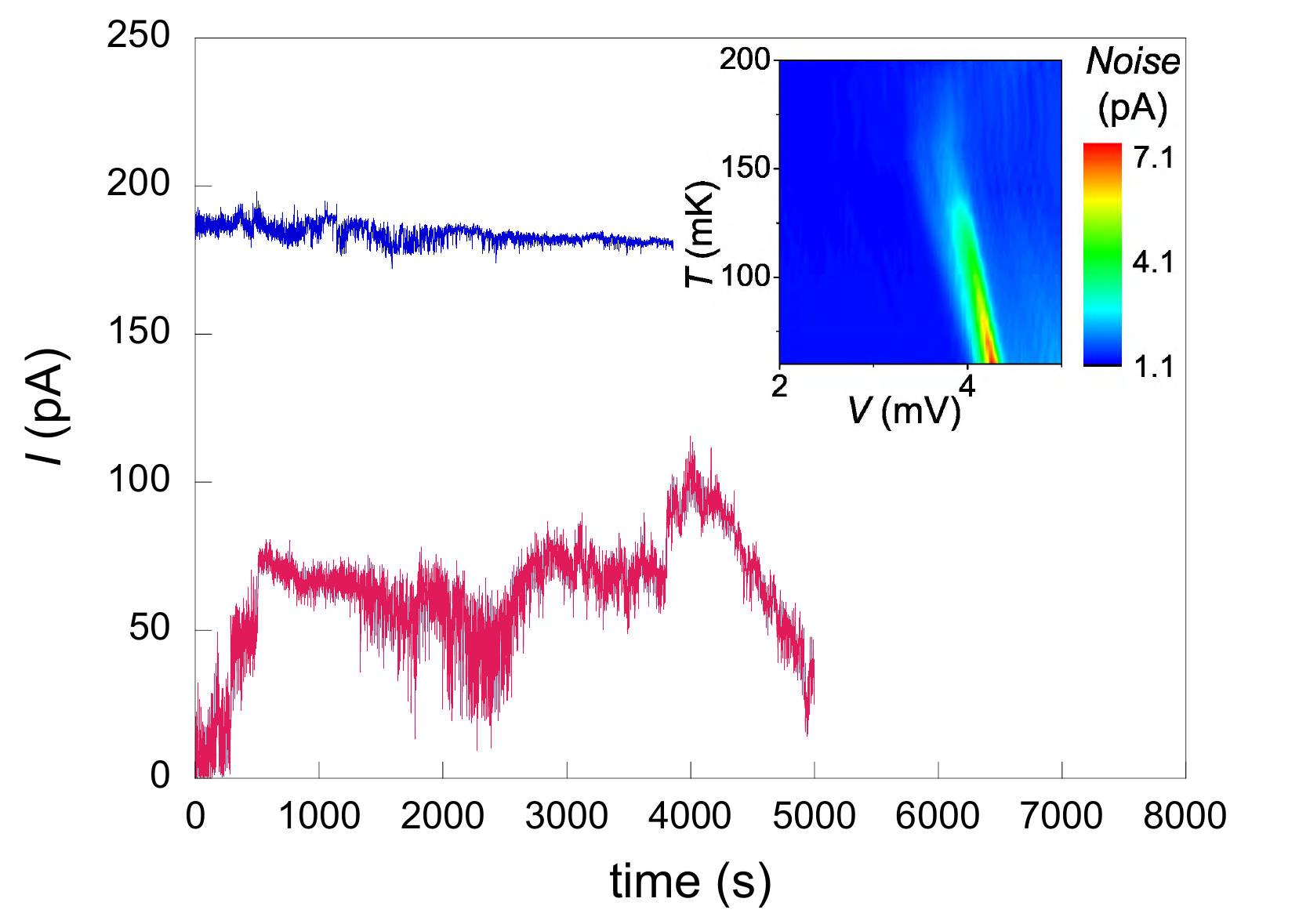}}
\caption{\label{fig2} The current as a function of time.  Current is plotted as a function of time for $n_{\text{s}}=5.2\times 10^{10}$~cm$^{-2}$ and $T=60$~mK at voltages $V=4.90$~mV (which lies between $V_{\text{th1}}$ and $V_{\text{th2}}$; lower curve) and $V=5.44$~mV (above $V_{\text{th2}}$). Inset: colour map of the broad-band noise at $n_{\text{s}}=5.36\times 10^{10}$~cm$^{-2}$ on a $(V,T)$ plane.}
\end{figure}

Figure~\ref{fig1} shows a set of low-temperature voltage-current curves at different electron densities in the insulating regime $n_{\text{s}}<n_{\text{c}}$ (here $n_{\text{c}}\approx 8\times 10^{10}$~cm$^{-2}$ is the critical density for the metal-insulator transition in this electron system); the corresponding interaction parameter given by the ratio of the Coulomb and Fermi energies, $r_{\text{s}}=g_{\text{v}}/(\pi n_{\text{s}})^{1/2}a_{\text{B}}$, exceeds $r_{\text{s}}\sim20$ (where $g_{\text{v}}=2$ is the valley degeneracy, $n_{\text{s}}$ is the areal density of electrons and $a_{\text{B}}$ is the effective Bohr radius in semiconductor). Two threshold voltages are observed at electron densities below $\approx 6\times 10^{10}$~cm$^{-2}$: with increasing applied voltage, the current is near zero up to a voltage threshold $V_{\text{th1}}$, then increases sharply until a second threshold voltage $V_{\text{th2}}$ is reached, above which the slope of the $V-I$ curve is significantly reduced and the behaviour is linear although not ohmic (see also the top inset to Fig.~\ref{fig1}). As the electron density is increased, the value of $V_{\text{th1}}$ decreases while the second threshold becomes less pronounced and eventually disappears. No hysteresis was observed for the range of electron densities studied. We point out that the observed behaviour (see also Fig.~\ref{fig2}) is quite distinct from that reported in the insulating state in amorphous InO films, where the current was found to jump at the threshold voltage by as much as five orders of magnitude and the $V-I$ curves exhibit hysteresis consistent with bistability and electron overheating\cite{ovadia09,altshuler09}.  It is important to note also that in our experiment the power dissipated near the onset $V_{\text{th1}}$ is less than $10^{-16}$~W, which is unlikely to cause significant electron overheating; for comparison, the power dissipated near the threshold voltage in Ref.\cite{ovadia09} is more than three orders of magnitude higher. Note also that the occurrence of the double threshold cannot be explained within the percolation picture in which case a single threshold is expected\cite{shashkin05}. Thus, the double threshold behaviour cannot be described within existing traditional models.

In the regime where both thresholds are present, the current measured at voltages between $V_{\text{th1}}$ and $V_{\text{th2}}$ exhibits strong fluctuations with time that are comparable to its value. Above the second threshold, $V_{\text{th2}}$, these anomalously large fluctuations disappear and the noise is barely perceptible. It is the sharp increase in noise on the $V-I$ curves that makes the two-threshold behaviour evident. The noise is shown explicitly in Fig.~\ref{fig2}, where the current is plotted as a function of time for density $n_{\text{s}}=5.2\times 10^{10}$~cm$^{-2}$.

Figure~\ref{fig3}(a) shows the $V-I$ characteristics for $n_{\text{s}}=5.36\times 10^{10}$~cm$^{-2}$ at different temperatures. As the temperature, $T$, is increased, the second threshold $V_{\text{th2}}$ becomes less pronounced and the threshold behaviour of the $V-I$ curves eventually smears out due to the shrinkage of the zero-current interval.

\subsection*{Noise.}
\begin{figure}
\scalebox{0.6}{\includegraphics{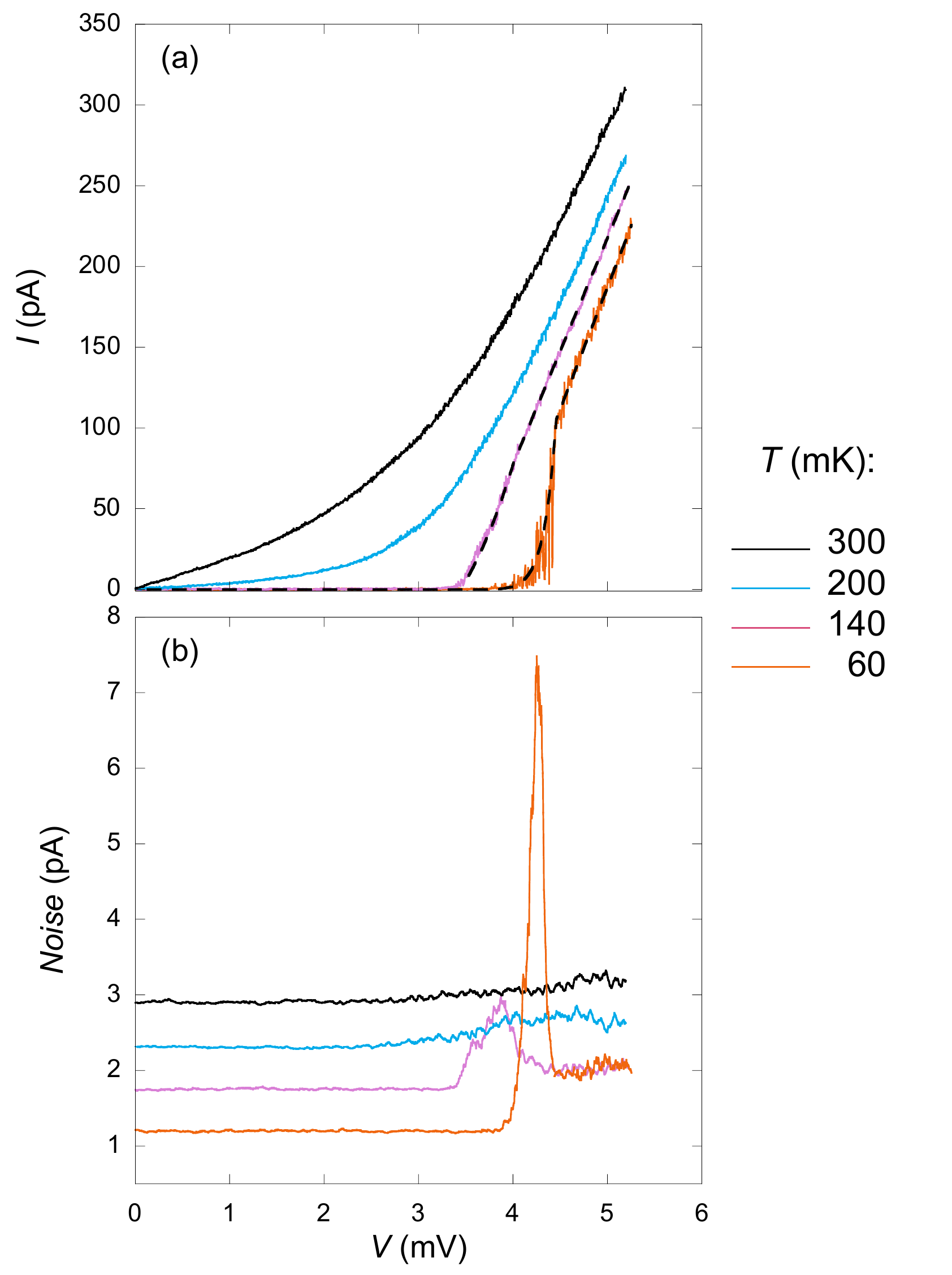}}
\caption{\label{fig3} Voltage-current characteristics and noise.  (a) $V-I$ characteristics at $n_{\text{s}}=5.36\times 10^{10}$~cm$^{-2}$ for different temperatures. The dashed lines are fits to the data using Eq.~(\ref{I}). (b)~The broad-band noise as a function of voltage for the same electron density and temperatures.  The three upper curves are shifted vertically for clarity.}
\end{figure}

The measured broad-band noise is shown as a function of voltage in Fig.~\ref{fig3}(b) for different temperatures at electron density $n_{\text{s}}=5.36\times 10^{10}$~cm$^{-2}$. The inset to Fig.~\ref{fig2} is a colour map of the broad-band noise on a $(V,T)$ plane. A large increase in the noise is observed between the thresholds $V_{\text{th1}}$ and $V_{\text{th2}}$ at the lowest temperature. This large noise decreases rapidly with increasing temperature in agreement with the two-threshold behaviour of the $V-I$ curves of Fig.~\ref{fig3}(a).

The spectrum of the generated noise, measured at its largest value, is displayed in Fig.~\ref{fig4}. The noise increases with decreasing frequency, $f$, according to the $1/f^{\alpha}$ law with $\alpha=0.6\pm0.1$, which is close to unity. This finding is consistent with the fact that the noise of the form $1/f^{\alpha}$ with $\alpha$ close to unity was found in the linear regime of response in similar samples near the metal-insulator transition\cite{jaroszynski02,jaroszynski04}. This indicates a universal behaviour of the noise spectrum.

\section*{Discussion}

We will now analyze our results in light of a phenomenological theory based on pinned elastic structures. There is a striking similarity between the double-threshold $V-I$ dependences in the low-density state of Si MOSFETs and those (with the voltage and current axes interchanged) known for the collective depinning of the vortex lattice in Type-II superconductors (see, \textit{e.g.}, Ref.\cite{blatter94}). The physics of the vortex lattice in Type-II superconductors, in which the existence of two thresholds is well known, can be adapted for the case of an electron solid. Current flows for zero voltage in a superconductor; the depinning of the vortex lattice occurs when a non-zero voltage appears. In our case, the situation is reciprocal: a voltage is applied but at first the current is zero in the limit of zero temperature; the depinning of the electron solid is signaled by the appearance of a non-zero current. The transient region between the dynamic ($V_{\text{d}}$) and static ($V_{\text{s}}$) thresholds corresponds to the collective pinning of the solid. In this region the pinning occurs at the centres with different energies and the current is thermally activated:
\begin{figure}\vspace{1.5mm}
\scalebox{0.45}{\includegraphics{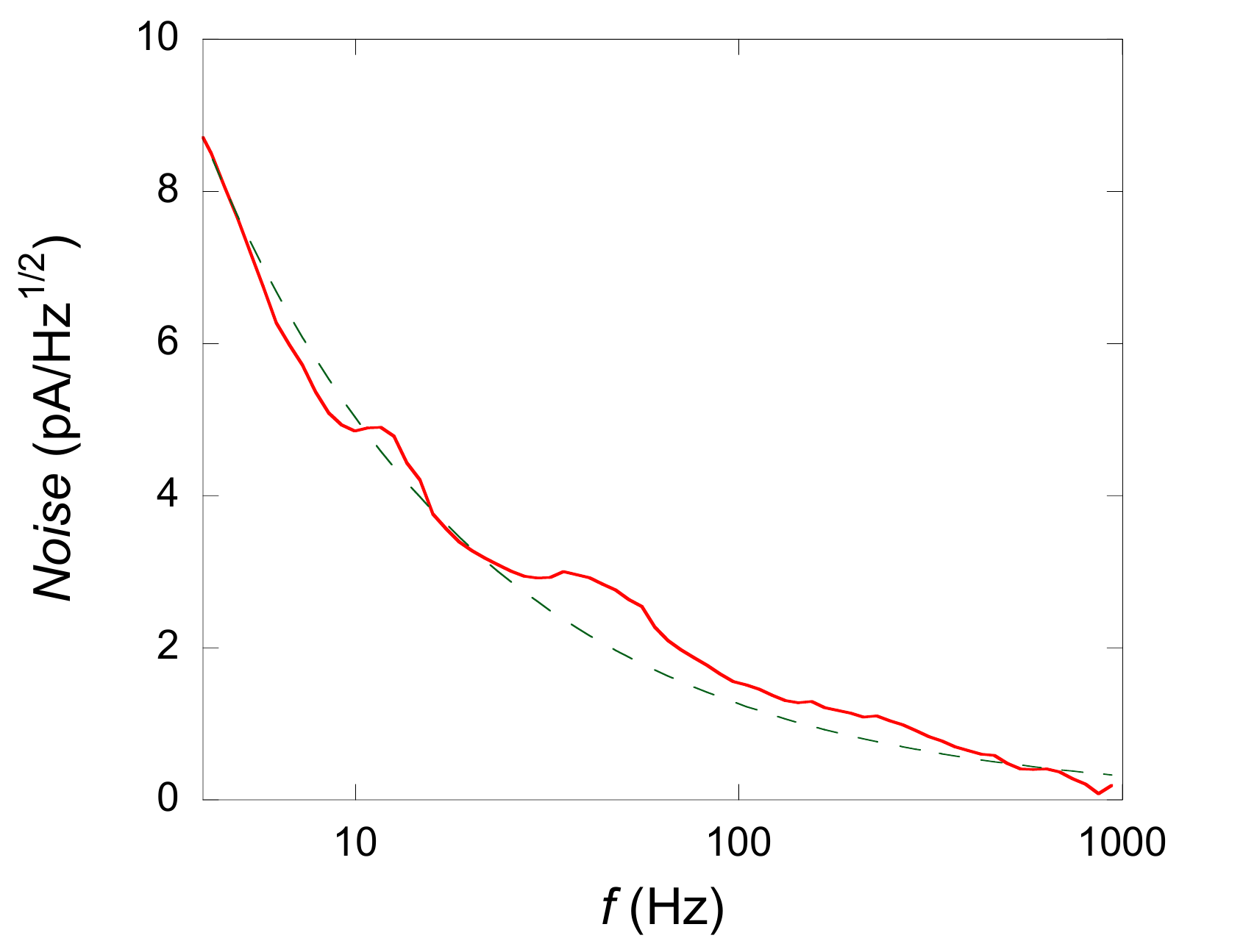}}
\caption{\label{fig4} Noise spectrum.  Noise is plotted as a function of frequency at $n_{\text{s}}=5.36\times 10^{10}$~cm$^{-2}$, $T=60$~mK and $V=4.26$~mV with resolution/bandwidth 2.5~Hz. The broad maxima at $f\sim10$ and 60~Hz on the order of $1$ pA~Hz$^{-1/2}$ are within the experimental uncertainty. The dashed line shows the $1/f^{0.6}$ dependence.}
\end{figure}
\begin{equation}
I\propto\exp\left[-\frac{U(V)}{k_{\text{B}}T}\right],
\label{exp}
\end{equation}
where $U(V)$ is the activation energy. The static threshold $V_{\text{s}}=V_{\text{th2}}$ signals the onset of the regime of solid motion with friction. This corresponds to the condition
\begin{equation}
eEL=U_{\text{c}},
\label{Uc}
\end{equation}
where $E$ is the electric field and $L$ is the characteristic distance between the pinning centres with maximal activation energy $U_{\text{c}}$. From the balance of the electric, pinning, and friction forces in the regime of solid motion with friction, one expects a linear $V-I$ characteristic that is offset by the threshold $V_{\text{d}}$ corresponding to the pinning force
\begin{equation}
I=\sigma_0(V-V_{\text{d}}),
\label{linear}
\end{equation}
where $\sigma_0$ is a coefficient; note that near $V_{\text{d}}$, one can in general expect a power-law behaviour of $(V-V_{\text{d}})$. Assuming that the activation energy for the electron solid is equal to
\begin{equation}
U(V)=U_{\text{c}}-eEL=U_{\text{c}}(1-V/V_{\text{s}}),
\label{U}
\end{equation}
we obtain the expression for the current
\begin{equation}
I=\left\{\begin{array}{c}
\sigma_0(V-V_{\text{d}}) {\text{ if }} V>V_{\text{s}}\\
\sigma_0(V-V_{\text{d}})\exp\left[-\frac{U_{\text{c}}(1-V/V_{\text{s}})}{k_{\text{B}}T}\right] {\text{ if }} V_{\text{d}}<V\leq V_{\text{s}}.
\end{array}\right.\label{I}
\end{equation}
The fits to the data using Eq.~(\ref{I}) are shown by dashed lines in Figs.~\ref{fig1} and \ref{fig3}(a). As seen from the figures, the experimental two-threshold $V-I$ characteristics are described well by Eq.~(\ref{I}). The value of $U_{\text{c}}$ decreases approximately linearly with electron density and tends to zero at $n_{\text{s}}\approx 6\times 10^{10}$~cm$^{-2}$ (the bottom inset to Fig.~\ref{fig1}). This is in contrast to the vanishing activation energy of electron-hole pairs at $n_{\text{c}}$ obtained by measurements of the resistance in the limit of zero voltages/currents\cite{shashkin01}. Presumably, the vanishing $U_{\text{c}}$ is related to the minimum number of the strong pinning centres for which the collective pinning is still possible. The fact that the coefficient $\sigma_0$ is approximately constant ($\sigma_0\approx 1.6\times 10^{-7}$~Ohm$^{-1}$) indicates that the solid motion with friction is controlled by weak pinning centres\cite{blatter94}. We argue that the large noise in the regime of the collective pinning of the solid between $V_{\text{d}}$ and $V_{\text{s}}$ should be suppressed in the regime of solid motion with friction at $V>V_{\text{s}}$. Indeed, in the regime of the collective pinning of the solid between $V_{\text{d}}$ and $V_{\text{s}}$, the solid deforms locally when the depinning occurs at some centre and then this process repeats at another centre \textit{etc}., which leads to the generation of a large noise. In contrast, in the regime of solid motion with friction above the second threshold $V_{\text{s}}$, the solid slides as a whole due to the overbarrier motion, resulting in the suppression of noise. Thus, the physics of pinned periodic/elastic objects is relevant for the low-density state in a 2D electron system in silicon MOSFETs.

One can estimate the characteristic range of frequencies of the generated noise $f\sim v_{\text{d}}/L$, where $v_{\text{d}}$ is the drift velocity. Using the parameters corresponding to the data shown in Fig.~\ref{fig3}(a) ($V_{\text{s}}\simeq4.46$~mV, $U_{\text{c}}/k_{\text{B}}\simeq1.7$~K, $I\sim20$~pA, $n_{\text{s}}=5.36\times 10^{10}$~cm$^{-2}$, and $L\sim10^{-3}$~cm as determined from Eq.~(\ref{Uc})), one obtains $f\sim500$~Hz, which is in reasonable agreement with the experiment. Note that in this experiment we did not observe the narrow band noise that is expected at frequencies close to the washboard frequency related to the motion of the solid (either the noise signal might be too small or the frequency of the noise might be too high). Although the model used describes the experiment successfully, further in-depth theoretical consideration is needed.

\section*{Methods}
\subsection*{Samples.} Measurements were made in an Oxford dilution refrigerator with a base temperature of $\approx50$~mK on (100)-silicon MOSFETs with a peak electron mobility close to 3~m$^2$V$^{-1}$s$^{-1}$ at $T<0.1$~K similar to those described in detail in Ref.\cite{heemskerk98}. The electron density was controlled by applying a positive dc voltage to the gate relative to the contacts; the oxide thickness was $150$~nm. Samples had a Hall bar geometry of width 50~$\mu$m. To overcome the main experimental obstacle in the low-density low-temperature limit --- high contact resistance --- thin gaps were introduced in the gate metallization that allow a high electron density to be maintained near the contacts regardless of the density in the main part of the sample. As a result, contact resistances did not exceed $\sim10$~kOhm and could be disregarded in the insulating state. 
\subsection*{\textit{V--I} characteristics and noise.} Measurements of the $V-I$ characteristics and noise were obtained in the main part of the sample with length 180~$\mu$m. The dc current and noise were measured using an ultra-low-noise current-voltage converter FEMTO DHPCA-100 (with 10~kOhm input resistance) connected to a digital voltmeter HP34401A Agilent Multimeter or lock-in SR830.  
\subsection*{Data availability.} The data that support the findings of this study are available from the corresponding author upon reasonable request.

\section*{Acknowledgments}
We are grateful to V.~T. Dolgopolov and A. Kapitulnik for pointing to the striking similarity between $V-I$ characteristics of the insulating state in Si MOSFETs and $I-V$ characteristics in Type-II superconductors. We are also indebted to Y.\ Gefen, P.\ Ghaemi, V. Kagalovsky, T.~M.\ Klapwijk, V.~V.\ Ryazanov, D.\ Shahar and V.~M.\ Vinokur for many illuminating discussions, to T.~M.\ Klapwijk for fabrication of the samples, and to Lucas Ho, Mike Pan, and Qing Zhang for technical help. This work was supported by NSF Grants DMR 1309337 and DMR 1309008, BSF Grant No.\ 2012210, RFBR 18-02-00368 and 16-02-00404, RAS, and the Russian Ministry of Sciences.
\end{document}